\pdfoutput=1
\def\myauthor{Panagiotis Agis Oikonomou-Filandras}
\documentclass[10pt, final, journal]{IEEEtran}
\usepackage{times,amsmath,epsfig,bm,amssymb}
\usepackage{algorithm}
\usepackage{graphicx}
\DeclareGraphicsExtensions{.pdf,.png,.jpg}
\usepackage{algorithmic}
\usepackage{epstopdf}
\usepackage{citesort}
\usepackage{color}

\def\mytitle{\LARGE Grid-Based Belief Propagation for Cooperative Localization\thanks{This work is supported under EPSRC grant EP/H011536/1.}}

\def\myauthor{Panagiotis-Agis Oikonomou-Filandras}

\def\supervisor{Kai-Kit Wong}

\def\citep{\cite}
\title{\mytitle}

\author{
{\myauthor, \supervisor, and Yangyang Zhang}\thanks{P. Oikonomou-Filandras and K. Wong are with the Department of Electronic and Electrical Engineering, University College London, WC1E 7JE, UK. Y. Zhang is with Kuang-Chi Institute of Advanced Technology, China.}
\vspace{1.6mm}\\
\fontsize{10}{10}\selectfont\itshape

\fontsize{9}{9}\selectfont\ttfamily\upshape
}
\begin{document}
\maketitle
\thispagestyle{empty}
\pagestyle{empty}

\def\bibliocommand{\bibliography{bibliography}}

\begin{thebibliography}{10}
\providecommand{\url}[1]{#1}
\csname url@samestyle\endcsname
\providecommand{\newblock}{\relax}
\providecommand{\bibinfo}[2]{#2}
\providecommand{\BIBentrySTDinterwordspacing}{\spaceskip=0pt\relax}
\providecommand{\BIBentryALTinterwordstretchfactor}{4}
\providecommand{\BIBentryALTinterwordspacing}{\spaceskip=\fontdimen2\font plus
\BIBentryALTinterwordstretchfactor\fontdimen3\font minus
  \fontdimen4\font\relax}
\providecommand{\BIBforeignlanguage}[2]{{%
\expandafter\ifx\csname l@#1\endcsname\relax
\typeout{** WARNING: IEEEtran.bst: No hyphenation pattern has been}%
\typeout{** loaded for the language `#1'. Using the pattern for}%
\typeout{** the default language instead.}%
\else
\language=\csname l@#1\endcsname
\fi
#2}}
\providecommand{\BIBdecl}{\relax}
\BIBdecl

\bibitem{Patwari:2005kc}
N.~Patwari \emph{et~al.}, ``Locating the nodes: Cooperative localization in wireless sensor networks,'' \emph{IEEE Signal Process. Mag.}, vol.~22, no.~4, pp. 54--69, Jul. 2005.

\bibitem{Wymeersch:2009hv}
H.~Wymeersch, J.~Lien, and M.~Win, ``Cooperative localization in wireless networks,'' \emph{Proc. IEEE}, vol.~97, no.~2, pp. 427--450, Feb 2009.



\bibitem{Ihler:2004jq}
A.~Ihler \emph{et~al.}, ``Nonparametric belief propagation for sensor self-calibration,'' in \emph{Proc. IEEE Int. Conf. Acoustics, Speech, and Signal Process.}, vol.~3, May 2004, pp. iii--861--4 vol.3.

\bibitem{priyantha2003anchor}
N.~B. Priyantha \emph{et~al.}, ``Anchor-free distributed localization in sensor networks,'' in \emph{Proc. Int. Conf. Embedded Networked Sensor Systems}, ACM, 2003, pp. 340--341.

\bibitem{shang2004improved}
Y.~Shang and W.~Ruml, ``Improved mds-based localization,'' in \emph{Proc. IEEE INFOCOM}, vol.~4, 2004, pp. 2640--2651.


\bibitem{ngamgrs}
``Military Map Reading 201'', [Online]. Available: http://http://earth-info.nga.mil/GandG/coordsys/mmr201.pdf

\bibitem{Buehrer:2010ew}
R.~Buehrer, T.~Jia, and B.~Thompson, ``Cooperative indoor position location using the parallel projection method,'' in \emph{Proc. Int. Conf. Indoor Positioning Indoor Navigation}, Sep. 2010, pp. 1--10.

\bibitem{oikonomou2011hybrid}
P.-A. Oikonomou-Filandras and K. K. Wong, ``Hybrid non-parametric belief propagation for localization in wireless networks,'' in \emph{Proc. Sensor Signal Process. for Defence}, Sep. 2011, pp. 1--5.

\bibitem{Yin:2015}
Yin, Feng, et al. ``Cooperative localization in WSNs using Gaussian mixture modeling: Distributed ECM algorithms.'',\emph{IEEE Trans. Signal Process.}, vol.~63, no.~6, pp. 1448--1463,  Mar 2015.


\bibitem{Koller:2009tn}
D.~Koller and N.~Friedman, \emph{Probabilistic graphical models: Principles and techniques}, Cambridge, MA : MIT Press, 2009.

\bibitem{Lien:2012bh}
J.~Lien \emph{et~al.}, ``A comparison of parametric and sample-based message representation in cooperative localization,'' \emph{Int. J. of Navigation and Observation}, vol. 2012, p.~10, 2012.

\bibitem{Caceres:2011wx}
M.~Caceres \emph{et~al.}, ``Hybrid cooperative positioning based on distributed belief propagation,'' \emph{IEEE J. Sel. Areas Commun.}, vol.~29, no.~10, pp. 1948--1958, Dec. 2011.

\bibitem{Ihler:2005be}
A.~Ihler \emph{et~al.}, ``Nonparametric belief propagation for self-localization of sensor networks,'' \emph{IEEE J. Sel. Areas Commun.},  vol.~23, no.~4, pp. 809--819, Apr. 2005.

\bibitem{Oliphant:2007dm}
T.~E. Oliphant, ``Python for scientific computing,'' \emph{Computing in Science \& Engineering}, vol.~9, no.~3, pp. 10--20, 2007.
\end{thebibliography}

\begin{abstract}
We present a novel parametric message representation for belief propagation (BP) that provides a novel grid-based way to address the cooperative localization problem in wireless networks. The proposed Grid-BP approach allows faster calculations than non-parametric  representations and works well with existing grid-based coordinate systems, e.g., NATO's military grid reference system (MGRS). This overcomes the hidden challenge inherent in all distributed localization algorithms that require a universally known global reference system (GCS), even though every node localizes using  arbitrary local coordinate systems (LCSs) for a reference. Simulation results demonstrate that Grid-BP achieves similar accuracy at much reduced complexity when compared to common techniques that assume {\em ideal} reference.
\end{abstract}

\begin{keywords}
Belief propagation, cooperative localization.
\end{keywords}

\vspace{-.1in}
\section{Introduction}\label{introduction}
The unprecedented adoption of mobile handhelds has created a host of new services that require accurate localization, even in GPS-denied environments. How to accurately localize without satellite positioning has been an active research topic. Various cooperative localization methods have been developed suitable to high noise scenarios, e.g., indoors \cite{Patwari:2005kc,Wymeersch:2009hv}.

\subsection{The Forgotten Challenge}\label{motivation}
The motivation behind this work is the assumption in the literature of distributed cooperative localization that requires a universally known {\em global coordinate system (GCS)}. Even in anchor-free algorithms, such as those in \cite{Ihler:2004jq,priyantha2003anchor,shang2004improved}, where nodes localize to a relative {\em local coordinate system (LCS)} only, the anchors with shared GCS knowledge are required to transform the local coordinates to global ones. Otherwise, nodes would have no idea the whereabout of the global origin. In the case of distributed ad-hoc networks, where the anchors are simply nodes with a good estimate of their GPS coordinates, achieving a shared GCS between them is non-trivial. Presumably, in this case, the anchors would have to communicate with each other to agree on a GCS with a common origin and pass that information to the rest of the network nodes. In addition, up-to-date information between nodes should be maintained, inducing an increased communication overhead.

The solution we suggest is to use directly a GCS for all calculations, therefore eliminating any need for LCS and all of the described issues. The first obvious choice of a GCS would be to use GPS as a common GCS to all nodes. Unfortunately using GPS coordinates in message passing operations would easily make calculations underflow due to the small distances inherent in indoors localization, and would require  scaling and normalization at each node. This means that every node would have to convert incoming messages to a LCS, suitable for message passing operations, and then convert them back to the GCS for transmission, increasing the computational cost of the algorithm. Instead, we propose a grid-based GCS that can be used directly to conduct message passing operations and at the same time hugely decrease the computational cost. In this manner we both remove any requirement for the networks to consent on a GCS and also achieve very low complexity.

\subsection{Our Contributions}
This correspondence proposes a novel scheme that uses a GCS system suitable for message passing algorithms in cooperative localization. As a real-life example, we use the NATO's military grid reference system (MGRS), cf. \cite{ngamgrs}.\footnote{Even though this letter uses the NATO's MGRS coordinate system, any grid-based coordinate system can be used with trivial changes.} The approach no longer requires GCS coordination between anchors. Also, it has inspired the use of parametric representations using multinomial probability mass functions (pmfs), allowing for a fast robust and accurate cooperative localization algorithm that elegantly resolves the GCS knowledge requirement. In summary, we have made the following contributions:
\begin{itemize}
\item We propose a grid-based GCS solution, i.e., map GPS coordinates to unique grid identifiers, solving the common reference issue in all distributed localization techniques.
\item Parametric approximations to the pmfs are proposed to overcome the computational bottleneck of non-parametric belief propagation (BP) used in cooperative localization.
\item Simulation results illustrate that the proposed grid-based BP method, which is referred to as Grid-BP, provides similar accuracy with low computational cost when compared to common techniques with {\em ideal} reference.
\end{itemize}

\section{Problem Formulation}\label{problemformulation}
We consider a network of nodes in a 2D environment which consists of $N$ agents and $M$ anchors, where $M\geq 4$ and $N \gg M$.  Let the space be subdivided into a square grid where each square ``bucket'' has a unique identifier, namely an ID. Then let $\bm X=[{X}_1,\dots,{X}_i,\dots,{X}_{N+M}] $ be the locations of all nodes,  with ${X}_{i}$ representing the unique identifier of node $i$ and $X_i \in \{x_1,x_2, \ldots x_k \}$, where $k$ iterates over all possible IDs. Also, let $\bm Z$ denote the coordinates of all nodes, with $\bm{Z}_{i}$ representing the coordinates of node $i$,  and the domain of $\bm{Z}_i$ is $\Re^2$.
The nodes communicate wirelessly and it is assumed that the maximum communication range for each node is $R_{\max}$. Time is slotted and time slots are denoted by the time index superscript $(t)$ for $t=1,2,\dots,\infty$. We represent the problem as a joint pmf. Let $p^{(t)}(X_{i})$ be the pmf, i.e., the belief that node $i$ has about its location at time $t$.  We model  $p^{(t)}(X_{i})$  as a multinomial distribution with parameters $\theta_k$, where $\theta_{k}$ is the probability of node $i$ being in ID $x_{k}$ and $\sum_{k}\theta_{k}=1$. In addition, let the set of all nodes $j$  within range of node $i$ be denoted as the  neighbourhood ${\cal N}_{i}$. Initially, the  belief for the agents can be a non-informative uniform pmf over the grid, while the anchors' pmfs are focused in the IDs close to the real position, e.g., within $10{\rm m}$. Anchors can obtain their IDs either by directly mapping them from satellite data, cf. \cite{ngamgrs}. Node $i$ receiving a message from node $j$ at time slot $t$ can derive, using time-of-arrival (ToA) measurements,\footnote{The assumption of using ToA is not restrictive on the proposed algorithm because it can easily be used with other measurement models.} a noisy estimate $r^{(t)}_{j\rightarrow i}$ of the distance between them. For convenience, we assume $r^{(t)}_{j\rightarrow i}=r^{(t)}_{i \rightarrow j}= r^{(t)}_{ji}$.

Thus, as in \cite{Buehrer:2010ew} for ToA distance measurements, we define the random variable $R_{ji}$ with its value $r_{ji}$ modelled as
\begin{equation}
r_{ji}=\| \bm{z}_i-\bm{z}_j\|+\eta_{ji},
\end{equation}
where $\eta_{ji}$ is a Gaussian noise with variance $\sigma^{2}_{ji}=K_{e}\|\bm{z}_i -\bm{z}_j\|^{\beta_{ji}}$ in which $K_{e}$ is a proportionality constant capturing the combined physical layer and receiver effect, and $\beta_{ji}$ denotes the path loss exponent. In the case of line-of-sight (LoS), $\eta_{ji}$ is assumed zero mean, and $\beta_{ji}=2$, i.e., $\eta_{ji} \sim \mathcal{N}(0, \sigma^{2}_{ji})$. In this work we assume only LoS, but it would be easy to extend the algorithm with NLoS mitigation, by using e.g. \cite{oikonomou2011hybrid,Yin:2015}.

We define the likelihood of node $i$ and node $j$ measuring distance $R_{ji}=r_{ji}$ between them at time $t$, given $X_i,X_j$ as
\begin{align}
p^{(t)}(R_{ji}=r_{ji}\mid X_i, X_j) \propto \exp\left(-\left(\frac{r_{ji}- ||C_i-C_j ||_2}{h}\right)^2\right)
\end{align}
where $h$ controls steepness, $C_i$ and $C_j$ are the coordinates of the centers of the grids' squares $X_i$ and $X_j$, respectively. Therefore, our objective is to find the maximum a posteriori (MAP), i.e., the values that maximize $p(\bm{X}|\bm{R})$ given distance measurements $\bm{R}=[R_{ji}]$. For a specific node $i$, we have
\begin{equation}
\hat{{X}}_{i} = \arg\max_{{X}_{i}}p^{(t)}({X}_i|\bm{R}_{i}).
\end{equation}
Thus, $p^{(t)}({X}_i|\bm{R}_i)$ can be evaluated using the Bayes' rule as
\begin{align}
p^{(t)}({X}_i|\bm{R}_{i})&\propto p^{(t)}({X}_{i}) \prod_{j \in {\cal N}_i} p^{(t)}(R_{ji}|{X}_{i})\notag\\
&\propto p^{(t)}({X}_{i}) \prod_{j \in {\cal N}_i}\int p^{(t)}(R_{ji}|{X}_i,{X}_j) p^{(t)}({X}_{j})\text{d}{X}_j,\label{eq:product_marginal}
\end{align}
in which the sign ``$\propto$'' means ``is proportional to'', and normalization should be done to obtain the pmf.

\section{The Proposed Grid-BP Algorithm}\label{mgrs-bpalgorithm}
Cooperative localization can be viewed as running inference on a probabilistic graphical model (PGM) \cite{Wymeersch:2009hv}, where messages are representations or probability functions. Here, the proposed grid-based localization algorithm will be presented. First, the use of a cluster graph and BP will be motivated, and the used PGM will be analysed. Then efficient approximation for the marginalization and the product operation will be given.

\vspace{-.1in}
\subsection{Belief Message Passing}\label{beliefmessagepassing}
\begin{figure}[!ht]
\centering
\includegraphics[scale=0.5]{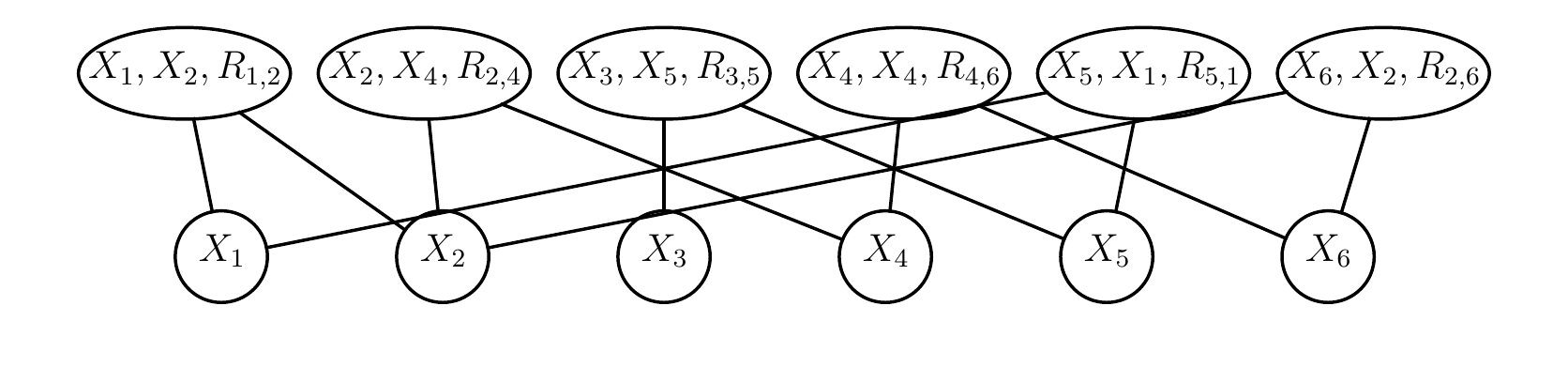}
\caption{The cluster graph. Lower row factors denote the node position beliefs. Upper row factors denote the ranging interactions between the nodes.}\label{fig:clustergraph}
\end{figure}

The network can be modelled as a cluster graph and loopy belief message passing algorithm can be used. We adopt a Bethe cluster graph \cite{Koller:2009tn}. The lower factors are composed of univariate potentials $\psi(X_{i})$. The upper region is composed of factors equal to $\psi( X_{i},X_{j}, R_{ji})$, e.g., see Fig.~\ref{fig:clustergraph}.

The lower factors are set to the initial beliefs for the given time slot $(t)$, and the upper factors to the corresponding cpmfs:

\begin{align}
\psi(X_{i})&=p^{(t)}(X_{i}),\\
\psi( X_{i},X_{j},R_{ji}= r^{(t)}_{ji})&= p^{(t)}(R_{ji}=r^{(t)}_{ji}|X_{i}, X_{j}).
\end{align}
Messages are then passed between nodes for multiple iterations until the node beliefs have converged. The message from node $j$ to node $i$, at BP iteration $(s+1)$ is calculated by
\begin{equation}\label{eq:messageupper}
\mu^{(s+1)}_{j \rightarrow i}(X_{i}) = \int \psi(X_{i},X_{j},R_{ji}=r^{(t)}_{ji}) \frac{b^{(s)}_{j\rightarrow i}(X_{j})}{\mu^{(s)}_{i\rightarrow j}(X_{j}) }\text{d} X_{j},
\end{equation}
where intuitively, a message \eqref{eq:messageupper} is the belief that node $j$ has about the location of node $i$  and $r^{(t)}_{ji}$ is the observed value of the distance between the nodes, at time slot $t$.

Then the belief of node $i$ is updated as
\begin{equation}
b^{(s+1)}_{i}(X_{i})= \lambda\psi(X_{i})\prod\limits_{k \in {\cal N}_{i}} \mu^{(s+1)}_{k \rightarrow i}(X_{i})
+ (1-\lambda)b^{(s)}_{i}(X_{i}), \label{eq:messagelower}
\end{equation}

where $\lambda$ is a dampening factor used to facilitate convergence.
BP continues until convergence, or if convergence is not guaranteed, $s$ reaches a maximum number of iterations $I_{\text{max}}$. Then the beliefs, representing approximations to the true marginals, for each node are found by \eqref{eq:messagelower}, i.e., $p^{(t+1)}(X_i)=b^{(s+1)}_{i}(X_{i})$. The proposed Grid-BP is given as Algorithm \ref{alg:HSM}. Each node needs to perform a marginalization operation \eqref{eq:messageupper}, and a product operation \eqref{eq:messagelower}. Approximations are required for both complex operations. In Grid-BP, we take advantage of the multinomial parametric form which we discuss next.

\begin{algorithm}
\caption{Grid-BP}
\begin{algorithmic}[1]
\label{alg:HSM}

\STATE Initialize beliefs $p^{(0)}(X_{i}) ~\forall i \in $ Nodes
\FOR{$t=0$ to $T$}
\FORALL{ $i \in$ Nodes}
\STATE Broadcast current belief $p^{(0)}(X_{i})$
\FORALL{$j \in  {\cal N}_{i}$ }
\STATE Collect distance estimates  $r^{(t)}_{ji}$
\ENDFOR
\ENDFOR
\STATE Initialize $\psi(X_{i}) =p^{(t)}(X_{i})$
\STATE Initialize $\psi(X_{i}, X_{j}, R_{ij})= p^{(t)}(R_{ij}= r^{(t)}_{ji} \mid X_{i}, X_{j})$
\REPEAT
\FORALL{ $i \in$ Nodes}
\FORALL{$j \in  {\cal N}_{i}$ }
\STATE Receive $b^{(s)}_{j}(X_{j}) $
\STATE Calculate  $\mu^{(s+1)}_{j \rightarrow i}(X_{i}) $, using \eqref{eq:messageupper} using Gibbs sampling (i.e., Algorithm \ref{alg:HSMgibbs})
\ENDFOR
\STATE Calculate  $b^{(s+1)}_{i}(X_{i})$, using \eqref{eq:messagelower}.
\STATE Check for convergence
\STATE Send $b^{(s+1)}_{i}(X_{i}) $
\ENDFOR
\UNTIL{convergence or $s$ reaches $I_{\text{max}}$}
\STATE Update belief $p^{(t+1)}(X_{i})$, using \eqref{eq:messagelower}.
\ENDFOR
\end{algorithmic}
\end{algorithm}
 \vspace{-.1in}

\subsection{Marginalization Operation \eqref{eq:messageupper}}\label{marginilizationoperation}
The calculation of \eqref{eq:messageupper} gives the belief node $j$ has about node $i$. To understand this, let us assume the case where all energy in  $b^{(s)}_{j}(X_{j}) $ is concentrated at a single ID $x_j$, then node $j$ would believe that  node $i$ is located in one of the IDs that approximate  a ``circle '' with centre $x_j$ and radius  $ r^{(t)}_{ji}$. Hence, to get $\mu^{(s+1)}_{j \rightarrow i}(X_{i}) $, first we draw $L$ particles from $x_j^{(l)} \sim b^{(s)}_{j}(X_{j}) $. Then we draw $L$ samples from  $\phi^{(l)} \sim {\cal U}(0, 2\pi)$ and $L$ samples from $\hat{r}^{(l)}_{ji}\sim \mathcal{N}(r^{(t)}_{ji}, h)$. The Gibbs sampling algorithm is provided as Algorithm \ref{alg:HSMgibbs}.  We repeat Algorithm  \ref{alg:HSMgibbs} for all incoming messages and we will get  $\{x_j^{(l)},\hat{r}_{ji}^{(l)},\phi^{(l)}\}_{l=1,j=1}^{L,\mid {\cal N}_i \mid}$.

\begin{algorithm}
\caption{Grid Gibbs Sampling}
\begin{algorithmic}[1]
\label{alg:HSMgibbs}
\STATE Set ${\cal D}_{ X_{i}}$ to empty
\FORALL{$j\in{\cal N}_{i}$}
\STATE Sample $ x^{(l)}_{j}\sim \mu_{j \rightarrow i}( X_{j})$ which is a multinomial pdf
\STATE Sample $\phi^{(l)} \sim {\cal U}[0, 2\pi]$
\STATE Sample $\hat{r}^{(l)}_{ji}\sim \mathcal{N}(r^{(t)}_{ji}, h)$
\STATE $ x^{(l)}_{i} ={\sf MAP}\text{-}{\sf DMtoID}(x^{(l)}_{j} , \hat{r}^{(l)}_{ji},\phi^{(l)})$ which maps the distance metric to IDs
\STATE Add $ \{  x^{(l)}_{i} \}_{l=1}^{L}$ to ${\cal D}_{ X_{i}}$
\ENDFOR
\RETURN ${\cal D}_{ X_{i}}$
\end{algorithmic}
\end{algorithm}

It is important to note that in order to combine the distance metric with the sampled IDs $\{x_j^{(l)}\}$ and get the set  ${\cal D}_i=\{x_i^{(l)}\}$, we use a mapping function which we define as
\begin{equation}
x_i^{(l)}={\sf MAP}\text{-}{\sf DMtoID}(x_j^{(l)},\hat{r}_{ji}^{(l)},\phi^{(l)})
\end{equation}
Intuitively we do this by counting for each sampled ID $x_j^{(l)}$ the number of IDs to the east and to the north, node $i$ will be, given the measured distance samples $\hat{r}_{ji}^{(l)}$ normalized by $D$ as
\begin{equation}\label{eq:steps}
\left[\begin{array}{c}
d_H\\
d_V
\end{array}\right]={\rm int}\left(\frac{\hat{r}_{ji}^{(l)}}{D}
\left[\begin{array}{c}
\cos(\phi^{(l)})\\
\sin(\phi^{(l)})
\end{array}\right]\right).
\end{equation}
Then we map the displacement $d_H, d_V$ to a new ID and return it as $x_i^{({l})}$. The set ${\cal D}_{ X_{i}}$ of all samples obtained from all incoming messages is used to find \eqref{eq:messagelower}. The distance to ID mapping function is given as Algorithm \ref{alg:MAP-DMtoID}. In the case of MGRS IDs the horizontal and vertical mappings are done by adding $d_H, d_V$ to the easting and northing components of the ID of $x_j^{({l})}$, to be discussed in Section \ref{militarygridreferencesystem}. There is no need to do any reverse mapping as we directly get the new IDs.

\begin{algorithm}
\caption{${\sf MAP}\text{-}{\sf DMtoID}$}
\begin{algorithmic}[1]
\label{alg:MAP-DMtoID}
\STATE Calculate horizontal and vertical steps using \eqref{eq:steps}
\STATE Map horizontal ID $ x^{(l)}_{j}\rightarrow  b^{(l)}_{j}$
\STATE  $b^{(l)}_{h}=b^{(l)}_{j} + d_H$
\STATE Inverse horizontal mapping $b^{(l)}_{h} \rightarrow x^{(l)}_{h}$
\STATE Map vertical ID $x^{(l)}_{h}\rightarrow b^{(l)}_{v}$
\STATE  $b^{(l)}_{i}=b^{(l)}_{v} + d_V$
\STATE Inverse vertical mapping  $b^{(l)}_{i} \rightarrow x^{(l)}_{i}$
\RETURN ${ x}^{(l)}_{i}$
\end{algorithmic}
\end{algorithm}

\subsection{Product Operation \eqref{eq:messagelower}}\label{productoperation}
To obtain \eqref{eq:messagelower}, we will first convert the samples calculated for each message  $\mu^{(s+1)}_{j \rightarrow i}(X_{i})$,  \eqref{eq:messageupper}, to parametric multinomial pmfs and then calculate their product. We assume that the parameters of each multinomial are random variables $\bm \Theta_i$  with a uniform Dirichlet prior with parameters $\bm \alpha =[\alpha_1 \cdots \alpha_K]$, where $K$ denotes the number of unique IDs in the pmf, i.e. $\mu^{(s+1)}_{j \rightarrow i}(X_{i})\approx p^{(s+1)}_{j \rightarrow i}(X_{i}|\bm \Theta_i; \bm \alpha  ) $.  Firstly we use the samples from each incoming message as observations and get the MAP estimate $ \hat{\bm\theta}_i=[\hat{\theta}_{i,1}\cdots \hat{\theta}_{i,K}]$ of the parameters $\bm \Theta_i$  of each $p^{(s+1)}_{j \rightarrow i}(X_{i}|\bm \Theta_i; \bm \alpha  ) $. We also assume that all incoming messages have the same prior distribution. This allows us to efficiently create parametric forms of the incoming messages as multinomial pmfs. These parameters are calculated as
\begin{equation}
\hat{\bm\theta_i}= {\mathbb E}\left[P(\bm\Theta_i | {\cal D}_{X_i})\right]= {\mathbb E} \left[p({\cal D}_{X_i}| \bm\Theta_i)p(\bm\Theta_i)\right],
\end{equation}
which gives
\begin{equation}\label{eq:likelihood}
\hat{\theta}_{i,k}=\frac{M_{k}+| {\cal N}_i | \alpha_{k}}{| {\cal N}_i |\sum\limits_{k}\left(M_{k}+\alpha_{k}\right)},
\end{equation}
where $M_{k}$ is number of particles $x_k $ in ${\cal D}_{X_i}$. The algorithm is presented in Algorithm \ref{alg:multiproduct}. For clarity, the quantities of each ID in the samples are shown as being found by a count function but in practice it can be done during the Gibbs sampling step allowing for a more efficient algorithm. Finally, with the parametric forms of each \eqref{eq:messageupper} in hand, we can easily calculate \eqref{eq:messagelower} as the dot product of the $\bm \theta_i$ parameters of all incoming messages and the node's own belief.

\begin{algorithm}
\caption{MAP Parameter Estimation}
\begin{algorithmic}[1]
\label{alg:multiproduct}
\STATE Let $|X_{i}|$ be the number of unique IDs in $p( X_{i}|\bm\Theta_{i})$
\STATE Calculate  $M_{k}={\rm count}( x_{k}, {\cal D}_{ X_{i}} ) ~\forall k \in | X_{i}|$
\FORALL {$k\in| X_{i}|$}
\STATE Calculate $\hat{\theta}_{i,k}$, with \eqref{eq:likelihood}
\ENDFOR
\RETURN$ {\hat{\bm \theta}_{i}} $
\end{algorithmic}
\end{algorithm}

\subsection{Message Filtering}\label{messagefiltering}
As it makes no sense to keep all the possible IDs on the planet wide GCS, we consider that each node constructs pmfs with the IDs only within a specific range $ R_{\text{grid}}$, e.g., within $100$m, of the IDs it receives in the first iteration. To reduce the IDs further, we propose a simple filter to only keep the most probable IDs summing up to an energy threshold. Simulation results (not included in this paper) suggest that by keeping $\sim80\%$ of the total energy of the pmf, the size of the messages is decreased by $\sim 90\%$ with no increase in localization error. Thus, assuming that each message covers a $100 \times 100{\rm m}^2$ grid, the total number of IDs used without the filter would be $10^4$. After the filter, only $\sim100$ IDs will be transmitted.

\subsection{Complexity}\label{complexity}
Most message passing cooperative localization algorithms tend to use a variant of Gibbs sampling to calculate \eqref{eq:messageupper} with similar computational costs \cite{Lien:2012bh}. Hence, the complexity cost tends to be bounded by the product operation \eqref{eq:messagelower} cost. As Grid-BP multiplies parametric form multinomials, the operation is bounded by $\bar{K} (| {\cal{N}}_{i} |+1)$, where $\bar{K}$ is the average number of IDs used in the calculations. The computational cost for the product operation of the compared algorithms is summarized in Table \ref{table:productcomplexity}, where $L$ is the number of particles  and $| {\cal{N}}_{i} |$ is the number of messages involved and $I_{\text{Hybrid-BP}}$ is the number of iterations the product algorithm in \cite{Caceres:2011wx} is run.

\begin{table}[H]
\caption{Comparison of complexity costs of \eqref{eq:messagelower}.$^\S$}
\center
\label{table:productcomplexity}

\begin{tabular}{c|c|c|c}
Approach & Algorithm & Complexity\\
\hline
Non-parametric & NBP  & $L_{\text{NBP}}^2(| {\cal{N}}_{i} |$+1)\\
Non-parametric & HEVA-BP  & $L_{\text{HEVA}}^2(| {\cal{N}}_{i} |$+1)\\
Parametric & Grid-BP  & $\bar{K}(| {\cal{N}}_{i} |$+1)\\
Parametric & Hybrid-BP  & $I_{\text{Hybrid-BP}} L_{\text{NBP}}| {\cal{N}}_{i} |(| {\cal{N}}_{i} |+1)$\\
\end{tabular}\\
$^\S$Note that $\bar{K}\simeq L_{\text{HEVA}}\ll L_{\text{NBP}}$, and typically $I_{\text{Hybrid-BP}}\approx 100$.
\end{table}

\section{Review of MGRS}\label{militarygridreferencesystem}
In this correspondence, as a grid based system, we employ MGRS, which is the geo-coordinate standard used by NATO military for locating points on the planet \cite{ngamgrs} and a combination of the universal transverse mercator (UTM) grid system and the universal polar stereographic (UPS) grid system, with a different labelling convention. It is essentially a global mesh grid that assigns a unique ID to each grid square. An example ID is ${\rm 10QCG12345678}$, where the first part ``${\rm 10Q}$'' is called the grid zone designator (GZD), the second part ``${\rm CG}$'' is the $100,000$-meter-square identifier, and finally the last numerical part gives the easting (first half digits) and northing (second half digits) inside the square identifier. Every two digits used (for a minimum of $2$ and a maximum of $10$) increase the resolution by a factor of $10{\rm m}$, down to a resolution of $1{\rm m}^2$ grid squares. Map coordinates are read from west to east first (easting), then from south to north (northing), i.e., left-right, down-up. In cases where the part of the ID is common to all neighbouring nodes, the common part can be dropped and only the rest need to be transmitted or used. For details on the specifics of MGRS, readers are referred to \cite{ngamgrs}.

\section{Simulation Results}

To assess the performance of Grid-BP, we conducted $300$ Monte-Carlo simulations and the root-mean-square (RMS) localization error was calculated for various noise levels.

In each simulation $100$ nodes with $20$ anchors are placed randomly in a $100{\rm m}\times100{\rm m}$ area and the communication range is limited to $12{\rm m}$ and $|{\mathcal N}_i|_{\text{avrg}}=4.03$. Anchor locations are modelled as multivariate Gaussian pdfs with an identity variance matrix. The grid resolution for Grid-BP is $D=1{\rm m}$. We compare Grid-BP with HEVA-BP \cite{oikonomou2011hybrid}, a computationally cheaper variation of non-parametric BP (NBP) \cite{Ihler:2005be}. We also compare it with a parametric belief propagation algorithm, namely Hybrid-BP \cite{Caceres:2011wx}.\footnote{For Hybrid-BP \cite{Caceres:2011wx}, we did not use information given from satellites.} In addition, the maximum number of message passing iterations is set to be $I_{\text{max}}=15$. The experiments were run for different noise levels with the noise factor used ranging between $K_e=[0.0 - 0.3]$ and the number of particles being $200$. Furthermore, to showcase the increased complexity of using GPS coordinates as a common GCS, a variant of HEVA that uses GPS was also provided. In HEVA-GPS, messages contain GPS coordinates that every node converts to an LCS before calculating \eqref{eq:messageupper} and \eqref{eq:messagelower}. Afterwards the updated beliefs are converted back to GPS coordinates and transmitted.

\begin{figure}[!ht]
  \centering
  \includegraphics[width=\columnwidth]{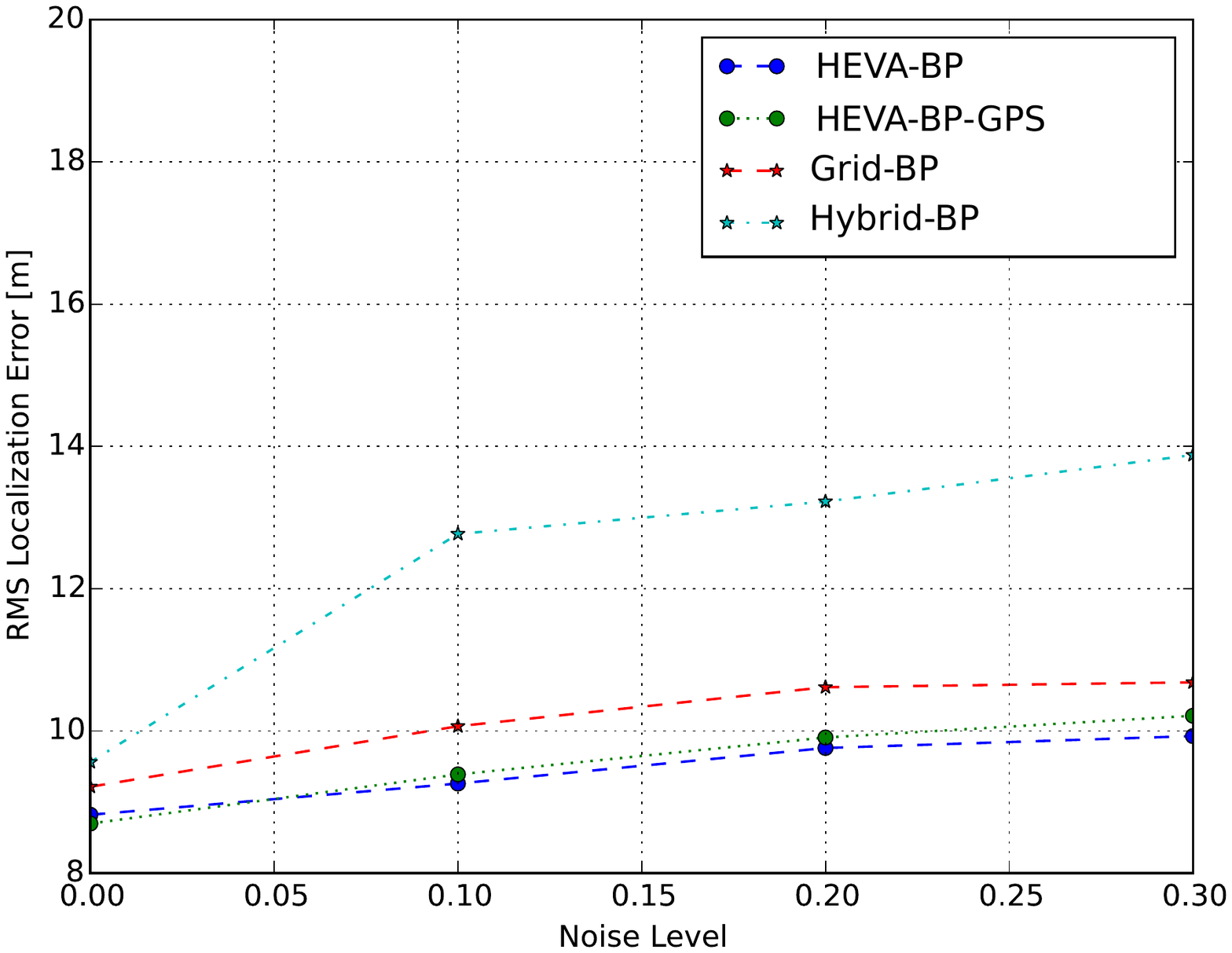}
  \caption{The rms error versus the amount of range estimation noise $K_{e}$ when the average node connectivity is $4.03$}\label{fig:noise}
\end{figure}

In Fig.~\ref{fig:noise}, the RMS localization error of all the algorithms as the noise coefficient increases is shown. We see that HEVA-BP and HEVA-BP-GPS have a better accuracy than both Grib-BP and Hybrid-BP. Grid-BP has a slightly higher RMS error.

\begin{figure}[!ht]
  \centering
\includegraphics[width=\columnwidth]{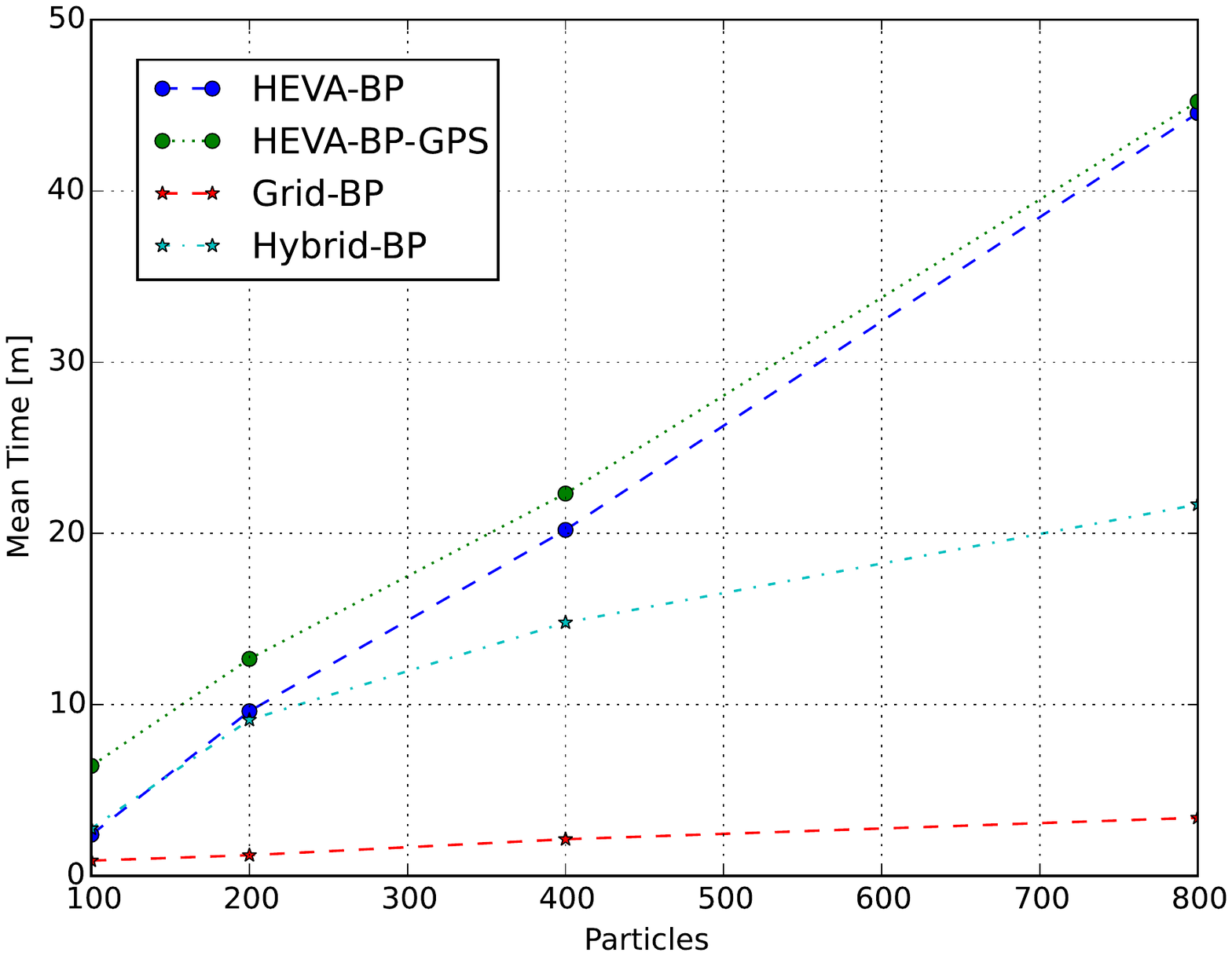}
\caption{The average time [m] versus the amount of particles used.}\label{fig:time}
\end{figure}

In Fig.~\ref{fig:time}, the average simulation time against the number of particles is shown. The improvement in computational cost by Grid-BP is observed, especially as the number of particles increases. It should also be noted that for higher particle numbers the cost of GPS scaling becomes almost insignificant compared to the cost of message passing equations.

Note that both algorithms, HEVA and Hybrid-BP have the strong assumption of sharing knowledge of the GCS origin, while Grid-BP and HEVA-GPS do not (the realistic scenario). Even though HEVA-BP-GPS performs as good as HEVA-BP, there is an increase in computational cost with HEVA-BP due to the mapping of the GPS coordinates to a local Cartesian reference frame (as can be observed in the mean simulation time in Fig.~\ref{fig:time}). As the number of particles gets higher, the relative computational efficiency of Grid-BP can also be seen. Note that all the simulations shown were run on an Intel i7 2.6GHz, using Python for scientific computing \cite{Oliphant:2007dm}.
\section{Conclusion}
This correspondence has proposed a novel parametric BP algorithm for cooperative localization that uses a grid-based system. The resulting Grid-BP algorithm combines a grid-based GCS that alleviates the hidden issue of requiring shared reference knowledge, and a parametric representation which allows quick and efficient inference. Simulation results showed that Grid-BP's performance is significantly better than other BP algorithms that rely on a known GCS. Grid-BP is also easy to be extended for mobile applications and add the non-LoS mitigation filter proposed in \cite{oikonomou2011hybrid}, making it a versatile and reliable choice in both military and civilian applications.


\end{document}